\documentclass[10pt, a4paper,twocolumn,aps,floatfix]{revtex4}
\usepackage{graphicx}
\usepackage{bm}

\begin{document}

\title{Evidence of secondary relaxations in the dielectric spectra of ionic
liquids}
\author{Alberto Rivera and Ernst A. R\"{o}ssler}
\affiliation{Experimentalphysik II, Universit{\"{a}}t Bayreuth, D 95440 Bayreuth, Germany}

\begin{abstract}
We investigated the dynamics of a series of room temperature ionic liquids
based on the same 1-butyl-3-methyl imidazolium cation and different anions
by means of broadband dielectric spectroscopy covering 15 decades in
frequency (10$^{-6}$-10$^{9}$Hz), and in the temperature range from 400 K
down to 35 K. An ionic conductivity is observed above the glass transition
temperature T$_{g}$ with a relaxation in the electric modulus
representation. Below T$_{g}$, two relaxation processes appear, with the
same features as the secondary relaxations typically observed in molecular
glasses. The activation energy of the secondary processes and their
dependence on the anion are different. The slower process shows the
characteristics of an intrinsic Johari-Goldstein relaxation, in particular
an activation energy \textit{E}$_{\beta }$\textit{=}24\textit{k}$_{B}$%
\textit{T}$_{g}$ is found, as observed in molecular glasses.
\end{abstract}

\maketitle

\vspace{1cm}

The glass transition involves a dramatic slowing down of the structural
relaxation in supercooled liquids from the ps time scale towards macroscopic
times which ultimately brings the liquid into the glassy state. The only
technique that can follow the evolution of the dynamics of a system in this
huge time scale window is dielectric spectroscopy, and studies covering the
full dynamic range (18 decades) of some paradigmatic glass formers are
available [1-3]. Dielectric spectroscopy showed that secondary relaxations
appear at frequencies higher than those of the main relaxation ($\alpha $%
-process), the importance of these processes being already pointed out more
than three decades ago by Johari and Goldstein (JG) [4]. They proposed that
these processes appear as a consequence of the glass formation, and they
demonstrated that one type of process, commonly termed now \textquotedblleft
JG $\beta $ relaxation\textquotedblright , occurs in supercooled liquids of
simple rigid molecules and therefore does not involve intramolecular motion.
When studying this intrinsic relaxation in different materials a similar
pattern shows up, with characteristic times showing an Arrhenius behavior
with an activation energy of 24\textit{k}$_{B}$\textit{T}$_{g}$ in most of
cases [5]. Yet the fundamental origin of secondary relaxations in
supercooled liquids is a matter of current attention and dispute, with much
effort being devoted to clarifying its true nature [6-9].

Secondary relaxations have been studied in many systems, but the search of
these processes in new types of materials should help to obtain a better
understanding of their origin. For example, it was reported recently that
binary systems show a continuous transformation between the high frequency
wing of the $\alpha $ relaxation and a JG relaxation when the composition of
the mixture is changed [10]. Ionic systems, formed by anions and cations
have been studied to a lesser extent. The most common ionic materials are
salts, like the NaCl, which have usually high melting temperatures. An
exception is calcium potassium nitrate Ca$_{\mathrm{0}\mathrm{.}\mathrm{4}}$K%
$_{\mathrm{0}\mathrm{.}\mathrm{6}}$(NO$_{\mathrm{3}}$)$_{\mathrm{1}\mathrm{.}%
\mathrm{4}}$ (CKN), with a rather low glass transition temperature \textit{T}%
$_{\mathrm{g}}$ at 333 K [11,12]. This molten salt was shown to exhibit a
secondary mechanical relaxation [13], but it was never clearly observed by
dielectric spectroscopy [4]. So the existence of secondary relaxations in
the dielectric spectra of purely ionic systems is an open question.

A new class of chemicals has been discovered in the last decades, the room
temperature ionic liquids (RTIL). They are molten salts formed by an anion
and a cation, like CKN, but the special choice of a bulky organic cation
makes them liquid down to unusually low temperatures, even below room
temperature [14]. Due to their ionic character, they show different
properties from the molecular liquids, specifically a wide range of
solubilities and non-measurable vapor pressure. Such interesting properties
promoted an intense research of these chemicals during the last decade as
perfect candidates for environmentally friendly or \textquotedblleft
green\textquotedblright\ chemistry, substituting the toxic organic solvents
used up to know in many industrial processes [15,16]. A great number of both
cations and anions can be used in the synthesis of RTIL offering high
versatility. For example, an unexpectedly large range of liquid fragilities
for these materials was reported in a detailed work on viscosity and dc
conductivity [17]. From all the different cations, the imidazolium
derivatives are among the most studied so far, since they offer the best
compromise of properties for their application [14-16]. The butyl chain
added to the imidazolium ring shows the minimum in the melting temperature,
and as a consequence, 1-butyl-3-methyl imidazolium (\textsc{bmim}) based
RTIL are easily supercooled [14-16]. For example, a quasielastic neutron
scattering study of \textsc{bmim} PF$_{\mathrm{6}}$ reported the existence
of a slow relaxation process showing the typical characteristics of the
glass transition dynamics ($\alpha $-relaxation) [18].

In this work we study the dynamics of a series of RTIL based on the same 
\textsc{bmim} cation but different anions: chloride (\textsc{cl}),
hexafluorophosphate (\textsc{pf}$_{\text{\textsc{6}}}$\textsc{),}
trifluoromethane sulfonate imide\textsc{\ (msf),} bis(trifluoromethane
sulphonate)imide\textsc{\ (bmsf)} in the temperature range from 400 K down
to 35 K. Differential scanning calorimetry (DSC) was used to detect phase
transitions of the samples and to extract the calorimetric glass transition
temperature \textit{T}$_{\mathrm{g}}$. The electric response of the RTIL was
probed by broadband dielectric measurements covering 15 decades (10$^{-6}$-10%
$^{9}$Hz). In this range ionic conductivity is observed above \textit{T}$_{%
\mathrm{g}}$ with a typical relaxation in the electric modulus
representation. Two secondary relaxations appear below \textit{T}$_{\mathrm{g%
}}$, with similar characteristics as the ones observed in molecular glasses.
One of them has all the features of an intrinsic JG relaxation in these
ionic systems.

The RTIL \textsc{bmim-cl, bmim-pf}$_{\text{\textsc{6}}}$\textsc{, bmim-msf,}
and \textsc{bmim-bmsf} ($\geq $97\%) were purchased from Sigma-Aldrich. In
order to probe the electric response, the samples as received were placed in
a stainless steel cell with an electrode separation of 39 $\mu $m. The high
surface over thickness ratio (S/d) of this cell resulted in an empty
capacitance of 48 pF. The electric response was measured at temperatures
from 35 K to 400 K and in the range 10 mHz - 10 MHz applying the gain-phase
analysis technique with a Solartron 1260 analyzer and a Novocontrol BDC-N
interface, placing the samples in an helium cryostat by Oxford. In the same
cryostat the characterization of samples was extended down to 35 K with the
Ultra Precision Capacitance Bridge Andeen-Hagerling AH2700, capable of
recording tan($\delta $)=$\varepsilon ^{\prime \prime }$/$\varepsilon
^{\prime }$ down to 10$^{\mathrm{-}\mathrm{6}}$ in the frequency range 50
Hz-20 kHz. The frequency range was extended to lower frequencies till 10$^{%
\mathrm{-}\mathrm{6}}$ Hz by using a time domain spectrometer with a
cryostat by Cryo Vac and recording the electric modulus relaxation in time
under helium gas atmosphere. The experimental setup to perform this
measurements as well as the Fourier transform procedure of the time domain
signal to get the spectral response were described elsewhere [19,20]. In the
frequency range 1 MHz-2 GHz an HP 4291B network analyzer was used, placing
the sample in a gold plated coaxial cell with 5 mm diameter and some 10 $\mu 
$m of electrode separation at the end of a transmission line. The
permittivity was automatically calculated from the reflection coefficients
of the line. This sample cell was placed inside an Oxford cryostat, and an
inert atmosphere of nitrogen was ensured during sample measurements.

\begin{figure}[h]
\includegraphics[width=1\columnwidth, viewport=12 20 250 180,
clip]{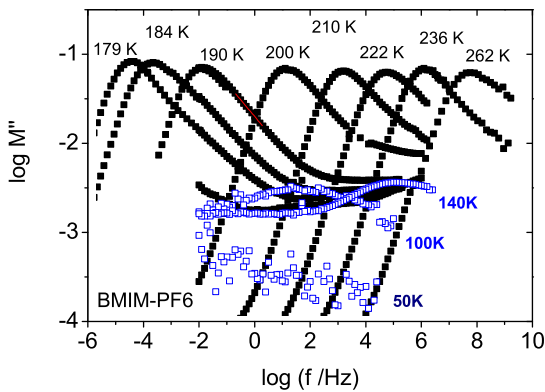}
\caption{Spectra of the imaginary part \textit{M}$^{\prime \prime }$\ of the
electric modulus of the ionic liquid \textsc{bmim-pf}$_{\text{\textsc{6}}}$
in the temperature range 50-262 K. Open symbols highlight the nearly
constant loss regime and a secondary relaxation. }
\end{figure}

DSC measurements show that the ionic liquids with \textsc{cl}, \textsc{pf}$_{%
\text{\textsc{6}}}$ and \textsc{bmsf} anions can be supercooled at the
cooling rate of 10 K/min, and they show a glass transition in the heating
curve at 208, 193 and 183 K respectively. In contrast, the sample containing
the \textsc{msf} anion shows a different pattern crystallizing at T$_{%
\mathrm{c}\mathrm{r}\mathrm{y}\mathrm{s}}$=241 K in the cooling experiment
even at higher cooling rates than 10 K/min.

\begin{figure}[h]
\includegraphics[width=1\columnwidth, viewport=12 20 250 180,
clip]{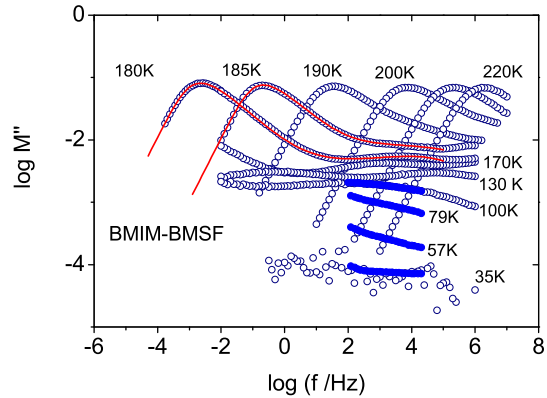} 
\caption{Spectra of the imaginary part of the electric modulus \textit{M}$%
^{\prime \prime }$\ for the ionic liquid \textsc{bmim-bmsf} in the
temperature range 35-220 K. Solid symbols are measured with an ultra
precision capacitance bridge. Lines are fits to a susceptibility function
including a secondary relaxation, used in molecular glasses (see text).}
\end{figure}

Figure 1 shows the electric modulus spectra \textit{M}$^{\prime \prime }$of
the \textsc{bmim-pf}$_{\text{\textsc{6}}}$ sample. These broadband
dielectric data covering 15 decades were obtained by overlapping
measurements of three different setups, a time domain, a frequency domain
and a high frequency network analyzer. The electric response of this RTIL is
representative for all the samples analyzed, and shows a typical ionic
conductor behavior [22, 23]. The ionic conduction appears as a relaxation in
the electric modulus representation \textit{M}$^{\ast }$\textit{=1/}$%
\varepsilon ^{\ast }$\textit{\ = i}$\omega \varepsilon _{\mathrm{0}}$\textit{%
/}$\sigma ^{\ast }$ [24], where the ionic conductivity process shows a step
in the real part \textit{M}$^{\prime }$, and a relaxation peak in the
imaginary part \textit{M}$^{\prime \prime }$\ (shown in figures 1 and 2).
The peak would have a symmetric Debye shape in the case of normal diffusion
(frequency independent conductivity curve), but due to the dispersion of the
conductivity curves of ionic conductors at high frequencies (sub-diffusion),
it presents an asymmetric maximum widened on the high frequency side. The
curves at 179 and 184 K show a somewhat distorted peak shape due to the
different cooling rates of the setups used for the time and frequency domain
measurements. The characteristic time of the conductivity relaxation can be
extracted from the frequency corresponding to the maximum value of the
imaginary part. When the temperature is reduced, the peak is shifted towards
lower frequencies. At frequencies higher than the peak, a nearly flat
response appears in our experimental frequency window, best seen in the 140
K curve of fig. 1 for 4 decades. The flat region of \textit{M}$^{\prime
\prime }$\ translate into a regime in which the imaginary part of the
permittivity, $\varepsilon ^{\prime \prime }$\textit{\ = M}$^{\prime \prime
} $\textit{/(M}$^{\prime }$\textit{+ M}$^{\prime \prime }$\textit{)}, stays
virtually constant, and therefore is usually called nearly constant loss
(NCL). This regime is universally found in ionic conductors [25,26]. At
still lower temperatures the NCL contribution is overshadowed by a broad
secondary process that shifts towards lower frequencies with decreasing
temperature. At the lowest temperature (50 K, figure 1), this relaxation
finally leaves our frequency window and the NCL contribution dominates again
the electric response. Thus, the NCL regime appears as a background
contribution, on top of which secondary relaxation processes are observed.

\begin{figure}[h]
\includegraphics[width=1\columnwidth, viewport=12 20 250 180,
clip]{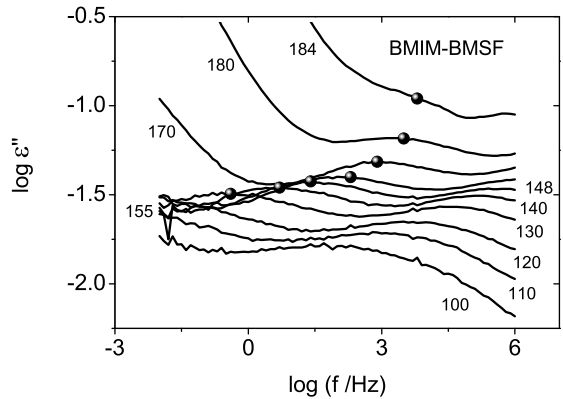} 
\caption{Detail of the spectra of the dielectric loss $\protect\varepsilon $%
\textquotedblright of \textsc{bmim-bmsf} at the temperatures that it shows
two secondary relaxations. Solid points indicate the evolution of the
relaxation amplitude approaching T$_{g}=$183 K. }
\end{figure}

In figure 2 the imaginary part of the spectra of the electric modulus of the
sample with the \textsc{bmsf} anion is plotted. Above \textit{T}$_{\mathrm{g}%
}$ the asymmetric conductivity relaxation peak is shifting in frequency with
temperature. Below \textit{T}$_{\mathrm{g}}$ the peak goes to very low
frequencies (mHz) and two weak secondary peaks are observed in the spectra.
They evolve with temperature and can be followed with the help of the ultra
precision bridge down to the lowest temperatures analyzed, 35 K, although in
a reduced frequency range (figure 2, solid circles). At the lowest
temperature also the secondary relaxation is gone and only a flat response
corresponding to the NCL regime is observed. The electric modulus spectra of
figs 1 and 2 are similar to the ones observed for the permittivity in
molecular glasses, with the main process being in our case the ionic
conductivity instead of the $\alpha $-process. This can be further tested
fitting the curves of figure 2 to susceptibility functions used for
molecular glasses. A Williams Watts ansatz, combining a generalized gamma
distribution for the main relaxation process with a modified Havriliak
Negami function compatible with a thermally activated process for the
secondary relaxation, proved it validity to fit the data of many glass
formers [7], and is used here to fit the modulus data. The lines in figure 2
show the good quality of the fits, with the difference to molecular liquids
that no high frequency wing is necessary to fit the data.

In the temperature range below T$_{\mathrm{g}}$ where we observed secondary
processes, the real part of the permittivity stays almost constant at a
value much higher than the imaginary part, thus the observed maximum in 
\textit{M}$^{\prime \prime }$\ translates into a maximum in $\varepsilon
^{\prime \prime }$, \textit{M}$^{\prime \prime }$= $\varepsilon ^{\prime
\prime }$/( $\varepsilon ^{\prime }$+ $\varepsilon ^{\prime \prime }$)
indicating that this maximum is not related to the long range conductivity.
The two secondary relaxations are clearly observed in the spectra of the
dielectric loss of the \textsc{bmsf} ionic liquid (figure 3). As a
consequence of the Kramers-Kronig relationships, both relaxations can be
identified as a step in the real part of the permittivity (figure 4). This
step or inflexion point translates to a maximum in the first derivative of
the curve. The inset of figure 4 is a plot of the derivative of the
permittivity data of the main figure, and each relaxation can be identified
as a peak. The points defining the maximum in the permittivity of figure 3
are showing that the amplitude of the slow process is increasing strongly
with the temperature at T$>$T$_{g}$ (top curve in fig. 3), as
observed in molecular glasses.

\begin{figure}[h]
\includegraphics[width=1\columnwidth, viewport=12 20 250 180,
clip]{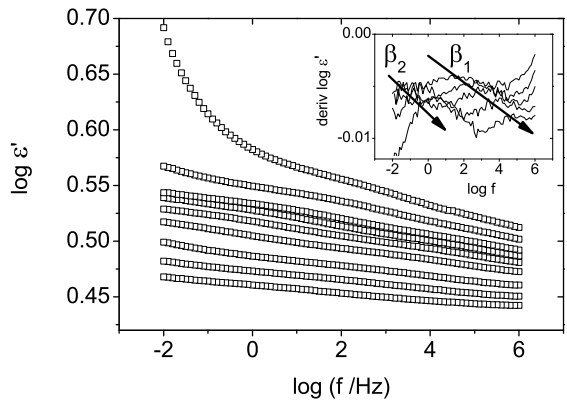} 
\caption{Frequency dependence of the real part of the permittivity $\protect%
\varepsilon $' in \textsc{bmim-bmsf} at tempeatures 100-180 K (from bottom
to top). The inset shows the derivative of the permitivitty spectra showing
the two relaxations as maxima (arrows). }
\end{figure}

In order to analyze the thermal behavior of these secondary relaxations, the
temperature dependence of the characteristic times $\tau =1/2\pi \upsilon
_{\max }$ is plotted in figure 5 in an Arrhenius representation, together
with the ones corresponding to the main relaxation process, the ionic
conductivity. While the $\tau $'s of the conductivity process exhibit a
non-Arrhenius behavior, all the secondary process show an Arrhenius linear
temperature dependence, with low activation energies of 0.38 and 0.22 eV
(lines in figure 5). Surprisingly, all the samples that are forming glasses
(the ones with the \textsc{cl, pf}$_{\text{\textsc{6}}}$ and\textsc{\ }%
\textsc{bmsf} anions) show a fast secondary process with not only the same
activation energy but also the same time constants, within experimental
error. This was not the case in the crystalline sample, which has the very
same cation, supporting the idea of Johari and Goldstein that beta
relaxations are intrinsic to the glassy state of matter [4]. The origin of
this fast relaxation can be tentatively related to the cation since it is
the element in common to all the samples. It was proposed in a neutron
scattering study of \textsc{bmim-pf}$_{\text{\textsc{6}}}$\ that the butyl
group is responsible for a relaxation in the liquid state [28], and indeed,
three minima are found in the potential energy of butane as a function of
the torsion angle, with a higher energy barrier of 20.5$\pm 0.4$ kJ/mol or
0.21$\pm 0.1$ eV [27]. The finding of the same activation energy within
experimental error for the fast relaxation as the energy barrier for the
change of conformations of butane agrees with this interpretation, but the
survival of conformational changes of the butyl group deep in the glass is
an open question, since the dielectric strength of the secondary process is
decreasing with temperature in the melt.

\begin{figure}[h]
\includegraphics[width=1\columnwidth, viewport=12 20 250 180,
clip]{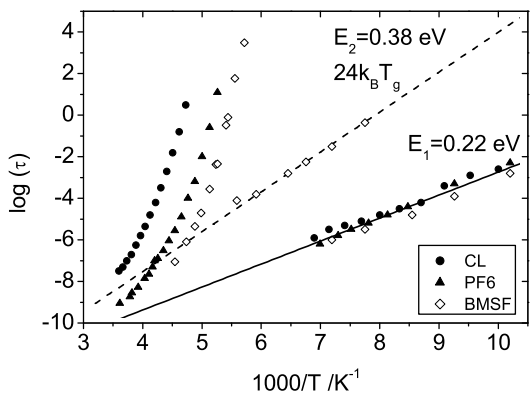}
\caption{Arrhenius plot of the correlation times of the main relaxations
process (at high temperatures) and the secondary relaxations of ionic
liquids with the same \textsc{bmim} cation and different anions detailed in
the legend. Dashed line shows the activation energy 24k$_{B}$T$_{g}$ for the 
\textsc{bmsf }sample.}
\end{figure}

The sample with the \textsc{bmsf} anion shows other secondary relaxation
with slower characteristic times (figure 5). Although it is only present in
this sample, and therefore its appearance is due to the presence of the 
\textsc{bmsf} anion, this slow process is merging with the main relaxation
(ionic conductivity). Moreover, it has an activation energy that correlates
well with the empirical relationship observed in many molecular glass
formers for a JG relaxation, \textit{E}$_{\mathrm{\beta }}$\textit{=}24%
\textit{\ k}$_{\mathrm{B}}$\textit{T}$_{\mathrm{g}}$[5]. Also the
pre-exponential factor of the temperature dependence of the characteristic
times, 10$^{\mathrm{-}\mathrm{1}\mathrm{4}}$-10$^{\mathrm{-}\mathrm{15}}$ s,
agrees with the ones usually observed in JG processes. We conclude that this
secondary relaxations is an intrinsic JG $\beta $ process in the ionic
liquid \textsc{bmim-bmsf}, and in the presence of long range Coulomb
interactions. Still, it is remarkable that this JG process is not found in
the other glass forming samples, that contain symmetric anions (\textsc{cl,
pf}$_{\text{\textsc{6}}}$) and therefore with no dipolar moment.

\bigskip The authors thank Alex Brodin for fruitful discussions.

\bigskip \textbf{REFERENCES}

[1] F. Kremer, A. Sch\"{o}nhals, \textit{Broadband Dielectric Spectroscopy}
(Springer, 2002).

[2] U. Schneider, P. Lunkenheimer, R. Brand, A. Loidl, Phys. Rev. E \textbf{%
59}, 6924 (1999).

[3] P. Lunkenheimer, U. Schneider, R. Brand, A. Loidl, Contemp. Phys. 
\textbf{41}, 15 (2000).

[4] G. P. Johari, M. Goldstein, J. Chem. Phys. \textbf{53}, 2372 (1970).

[5] A. Kudlik, S. Benkhof, T. Blochowicz, C. Tschirwitz, and E. R\"{o}ssler,
J. Mol. Struct. \textbf{479}, 210 (1999).

[6] U. Schneider, R. Brand, P. Lunkenheimer, and A. Loidl, Phys. Rev. Lett. 
\textbf{84}, 5560 (2000).

[7] T. Blochowicz, C. Tschirwitz, S. Benkhof, and E. A. R\"{o}ssler, J.
Chem. Phys. \textbf{118}, 7544 (2003).

[8] K. Duvvuri and R. Richert, J. Chem. Phys. \textbf{118}, 1356 (2003).

[9] K. L. Ngai and M. Paluch, J. Chem. Phys. \textbf{120}, 857 (2004).

[10] T. Blochowicz and E. A. R\"{o}ssler, Phys. Rev. Lett. \textbf{92},
225701 (2004).

[11] S. Sen and J. F. Stebbins, Phys. Rev. Lett. \textbf{78}, 3495 (1997).

[12] P. Lunkenheimer, A. Pimenov, and A. Loidl, Phys. Rev. Lett. \textbf{78}%
, 2995 (1997).

[13] C. Mai, S. Etienne, J. Perez and G. P. Johari, Philos. Mag. B \textbf{50%
}, 657 (1985).

[14] P. Wasserscheid and T. Welton, \textit{Ionic Liquids in Synthesis}
(Wiley, Weinheim, 2003).

[15] T. Welton, Chem. Rev. (Washington, D.C.) \textbf{99}, 2071 (1999).

[16] P. Wasserscheid and W. Keim, Angew. Chem., Int. Ed. Engl. \textbf{39},
3772 (2000).

[17] W. Xu, E. I. Cooper, and C. A. Angell, J. Phys. Chem. B \textbf{107},
6170 (2003).

[18] A. Triolo, O. Russina, F. Juranyi, S. Janssen, and C. M. Gordon, J.
Chem. Phys. \textbf{119}, 8549 (2003).

[19] H. Wagner and R. Richert, J. Appl. Phys. \textbf{85}, 1750 (1999).

[20] A. Rivera, T. Blochowicz, C. Gainaru, and E. A. R\"{o}ssler, J. Appl.
Phys. \textbf{96}, 5607 (2004).

[21] T. Blochowicz, E. A. R\"{o}ssler, J. Chem. Phys. \textbf{122}, 224511
(2005).

[22] A. K. Jonscher, \textit{Dielectric Relaxation in Solids} (Chelsea
Dielectric Press, London, 1983).

[23] J. C. Dyre, T. B. Schroeder, Rev. Mod. Phys. \textbf{72}, 873 (2000)$.$

[24] P. B. Macedo, C. T. Moynihan, and R. Bose, Phys. Chem. Glasses 13, 171
(1972).

[25] W. K. Lee, J. F. Liu, A. S. Nowick, Phys. Rev. Lett. 67, 1559 (1991).

[26] C. Leon , A. Rivera, A. Varez, J. Sanz, J. Santamaria, K. L. Ngai,
Phys. Rev. Lett. \textbf{86}, 1279 (2001).

[27] N. L. Allinger, R. S. Grev, B. F. Yates, H. F. Schaefer III, J. Am.
Chem. Soc. \textbf{112}, 114 (1990).

[28] A. Triolo, O. Russina, C. Hardacre, M. Nieuwenhuyzen, M. A. Gonzalez,
H. Grimm, J. Phys. Chem. B, in press.

\end{document}